\documentclass[twocolumn,superscriptaddress,showpacs,amsfonts]{revtex4-1}
\pdfoutput=1

\usepackage{epsfig}
\usepackage{graphicx}
\usepackage{amsmath}
\usepackage{amssymb}
\usepackage{color}
\usepackage{bm}

\begin{document}
\title{Designing Electron Spin Textures and Spin Interferometers by Shape Deformations} 
\author{Zu-Jian Ying}
\affiliation{CNR-SPIN and Dipartimento di Fisica ``E. R. Caianiello'',
Universit\`a di Salerno, I-84084 Fisciano (Salerno), Italy}
\affiliation{Beijing Computational Science Research Center, Beijing 100084, China}
\author{Paola Gentile}
\affiliation{CNR-SPIN and Dipartimento di Fisica ``E. R. Caianiello'',
Universit\`a di Salerno, I-84084 Fisciano (Salerno), Italy}
\author{Carmine Ortix}
\affiliation{Institute for Theoretical Solid State Physics, IFW-Dresden, Helmholtzstr. 20, D-01069 Dresden, Germany}
\affiliation{Institute for Theoretical Physics, Center for Extreme Matter and Emergent Phenomena, Utrecht University, Leuvenlaan 4, 3584 CE Utrecht, The Netherlands}
\author{Mario Cuoco}
\affiliation{CNR-SPIN and Dipartimento di Fisica ``E. R. Caianiello'',
Universit\`a di Salerno, I-84084 Fisciano (Salerno), Italy}

\begin{abstract}
We demonstrate that the spin orientation 
of an electron propagating in a one-dimensional nanostructure with Rashba spin-orbit (SO) coupling 
can be manipulated on demand by changing the geometry of the nanosystem.
Shape deformations that result in a non-uniform curvature give rise to complex three-dimensional spin textures in space.
We employ the paradigmatic example of an elliptically deformed quantum ring to unveil the way to get 
an all-geometrical and all-electrical control of the spin orientation.
The resulting spin textures exhibit a tunable topological character with windings around
the radial and the out-of-plane directions. 
We show that these topologically non trivial spin patterns 
affect the spin interference effect in the deformed ring, thereby resulting in different geometry-driven ballistic electronic transport behaviors.
Our results establish a deep connection between electronic spin textures, spin transport and the nanoscale shape of the system. 
\end{abstract}
\maketitle
\noindent 
{\it{Introduction -- }} 
The manipulation and control of the electron spin are essential ingredients for the development of 
innovative quantum-engineered devices~\cite{Awschalom2013,Fabian2004,Chappert2007}.
The spin-orbit (SO) coupling
~\cite{Dresselhaus1955,Rashba1960,Bychkov1984} 
is particularly attractive in this framework
because it offers the promising prospect~\cite{Manchon2015} of an all-electrical intrinsic control over the spin without applying a magnetic field.
For low-dimensional nanosystems with structure inversion asymmetry, 
this coupling between the orbital motion and the electron spin is due to 
the so-called Rashba SO interaction ~\cite{Dresselhaus1955,Rashba1960,Bychkov1984}. 
The Rashba SO is at the heart of a growing research interest for the spin generation, manipulation and detection 
due to the tantalizing possibilities of tuning
the spin orientation through the electron propagation and vice versa to exert a spin control of the electron trajectories.
Central consequences of these constituent features are, among the many possibilities, the Spin Hall~\cite{spinhall1,spinhall2} and the spin 
galvanic effects~\cite{spingalvanic1,spingalvanic2,spingalvanic3}, the SO driven spin-torque~\cite{spinorbittorque1,spinorbittorque2}, 
the design of quantum topological states~\cite{Hasan}, etc.

Charge carriers in materials with Rashba SO are subject to a momentum dependent effective magnetic field which not only results into a spin-dependent velocity but also
in a non trivial geometric phase. Such phase is another important mean to manipulate the electron spin. 
For instance, in the Aharonov-Casher effect~\cite{ACeffect}, magnetic dipoles moving around a tube
of electric charge acquire a non trivial Berry~\cite{Berry} or geometric phase~\cite{AAphase}, being dual to the case
of the Aharonov-Bohm~\cite{ABeffect} effect for charged
particles moving in a closed circuit around a magnetic flux.
With the development of semiconducting
nanostructures, a possibility has emerged to tune quantum states by combining 
Rashba effects and geometric phases. Considering the role of magnetic field textures in driving a spin geometric phase~\cite{Loss1990}, 
the effects of quantum geometric phases have been predicted~\cite{Frustaglia2004} and experimentally observed in the transport properties 
of semiconducting quantum rings~\cite{Nagasawa2012,Nagasawa2013}. 
Such findings opened the path to a spin topological design~\cite{Lyanda-Geller1993}, as demonstrated by the magnetic field driven
topological transition in the geometric phase and the consequences on the spin-transport~\cite{Nagasawa2013,Saarikoski2015}. 
 
In view of the rapid progress in nanostructuring techniques, it would be highly challenging and desirable to design nanostructures 
where the system geometry is the main knob to tailor
the electron spin properties. In this Letter, we demonstrate that the curvature of a 
nanostructure with Rashba SO coupling allows to
manipulate and control the spin orientation and 
the spin transport. We use the prototypical geometry of elliptical quantum rings to unveil 
complex three-dimensional spin textures in space and how they evolve in the presence of a non uniform
curvature. The resulting spin patterns have distinct topological features with tunable windings around
the radial and the out-of-plane spin directions along the elliptical ring. Remarkably, 
we find different types of transitions between quantum states with inequivalent spin textures
by varying the curvature strength or the Rashba SO coupling thus
indicating a unique path to control the electron spin in curved nanostructures.
We then demonstrate a fundamental twist between curvature, spin texture and spin transport: 
it is the topology of the spin texture with non trivial windings that can drive an all-geometric changeover 
of the conductance of the loop.
\begin{figure}
\includegraphics[width=0.95\columnwidth]{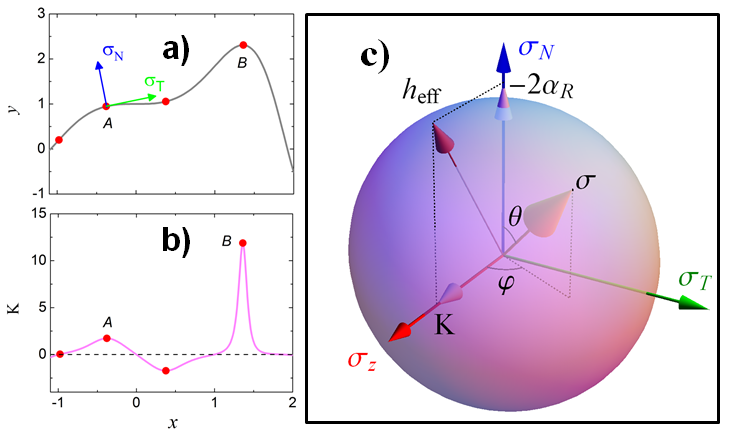}
\caption{Schematics of (a) the profile of a planarly curved nanowire and (b) the related curvature $K(s)$. 
(c) denotes the Bloch sphere in the moving frame of the electrons with the vectors associated to the electron spin orientation and the effective field $\boldsymbol{h}_{\text{eff}}$.}
\label{fig:fig1}
\end{figure}   

{\it{Curvature-driven spin torques--}}
Electrons confined to move along a one-dimensional planarly curved nanostructure [Fig. \ref{fig:fig1}(a)] are subjected to a Rashba SO interaction, 
due to the inversion symmetry breaking, 
that, as for the case of a straight nanostructure, couples the orbital momentum and the local spin component that is normal to the electron motion~\cite{Gentile2013}. 
Since the nanostructure has a non trivial curvature, the local normal spin direction depends
on the position and thus generally it does not commute with the momentum. Such interplay  
can be conveniently expressed by introducing the local normal $\hat{\cal N}(s)$ and tangential $\hat{\cal T}(s)$
directions at a given position $s$ along the curve, as well as the related local Pauli matrices
$\sigma_N(s) = \boldsymbol{\tau} \cdot \hat{\cal N}(s)$ and $\sigma_T(s) = \boldsymbol{\tau} \cdot \hat{\cal T}(s)$ in the moving frame of the electrons,
$\boldsymbol{\tau}$ being the usual Pauli matrices [Fig. \ref{fig:fig1}(a)].
Then, the Hamiltonian for a planarly curved nanostructure with Rashba SO can be written as ~\cite{Gentile2013,Ortix2015,Gentile2015}:
\begin{equation}
{\cal H}_{\bf k \cdot p}= -\dfrac{\hbar^2}{2 m^{\star}} \partial_s^2 + \dfrac{i \alpha_{SO}}{2} \left[\sigma_N(s) \partial_s + \partial_s \sigma_N(s)\right],
\label{eq:hamiltoniankp}
\end{equation}
where $s$ is the arclength of the planar curve measured from an arbitrary reference point, $m^{\star}$ is the effective mass of the charge carriers, and 
$\alpha_{SO}$ is the Rashba SO coupling strength. 
It is convenient to express the normal and tangential directions to the curve in terms of a polar angle
$f(s)$ as $\hat{\cal N}(s) = \left\{\cos{f(s)},\sin{f(s)} ,0 \right\}$, and  $\hat{\cal T}(s) = \left\{\sin{f(s)} , -\cos{f(s)},0 \right\}$. Using 
the Frenet-Serret type equation of motion~\cite{refA1}, $\partial_s \hat{\cal N}(s) = K(s) \hat{\cal T}(s)$, where  $K(s)$ is the local curvature, 
we can immediately relate the polar angle to the local curvature via  $\partial_s f(s) = -K(s)$ [Figs. \ref{fig:fig1}(a,b)]. 
To proceed further, we use the fact that a spin eigenmode $|\Psi _{E}\rangle$ of the Hamiltonian in Eq.~\ref{eq:hamiltoniankp} evolves in space according to 
$i \partial_s |\Psi _{E}\rangle= G(s) |\Psi _{E}\rangle \label{evolve}$ where $G(s)=- \left( \alpha_R \sigma_N(s)+ c_0\,\sigma_0\right)$,  
$\alpha_R=\frac{m^{*}\alpha_{SO}}{\hbar^2}$ being the inverse Rashba SO length, $c_0^2=\frac{2 m^{*} E}{\hbar^2}+\alpha_{R}^2$ relates to the eigenergy $E$.  
This relation, in turn, allows us to derive the spin-torque exerted on the electron spin through a fundamental equation 
that links the geometric curvature of the nanostructure, the Rashba SO coupling and the electron 
spin orientation in the Frenet-Serret frame~\cite{SupMat}:  
\begin{equation}
\partial_s \langle \boldsymbol{\sigma} \rangle = \boldsymbol{h}_{\text{eff}} \times \langle \boldsymbol{\sigma} \rangle \,. \label{eq:gyroscope}
\end{equation}
Here, the local spin orientation is $\langle \boldsymbol{\sigma} \rangle=\{\langle {\sigma}_T \rangle,\langle {\sigma}_N \rangle,\langle {\sigma}_z \rangle \}$ while the 
effective spin-orbit field $\boldsymbol{h}_{\text{eff}}=\{0, -2\alpha_R, K(s) \}$ lies in the normal-binormal plane and depends on the local curvature and $\alpha_R$. 
The spin direction then defines a Frenet-Serret-Bloch sphere which is expressed in terms of the azimuthal and polar angles $\theta(s)$ and $\varphi(s)$ [Fig. \ref{fig:fig1}(c)]. 
Eq.~\ref{eq:gyroscope} generally implies that due to a non zero curvature, the electron spin acquires a finite out-of-plane binormal $\hat{z}$ component. More importantly, 
a non trivial component along the tangential direction appears provided the curvature is not constant. Albeit the derivative $\partial_s$ of the spin vector locally vanishes 
if the spin is aligned to the effective spin-orbit field, the variation of the local curvature yields a non-vanishing torque which results into a component of the spin vector parallel to
the electron propagation direction. 
Insights into the electron spin trajectories regulated by Eq.~\ref{eq:gyroscope} can be gained by looking at the spin velocity vector field 
around the point on the Frenet-Serret-Bloch sphere indicated by $\boldsymbol{h}_{\text{eff}}$~\cite{SupMat}. It exhibits a vortex-like 
structure whose dynamical behavior reveals the role of key geometrical control parameters as the first and second derivatives
of the curvature $K(s)$~\cite{SupMat}.
\begin{figure}
\includegraphics[width=0.95\columnwidth]{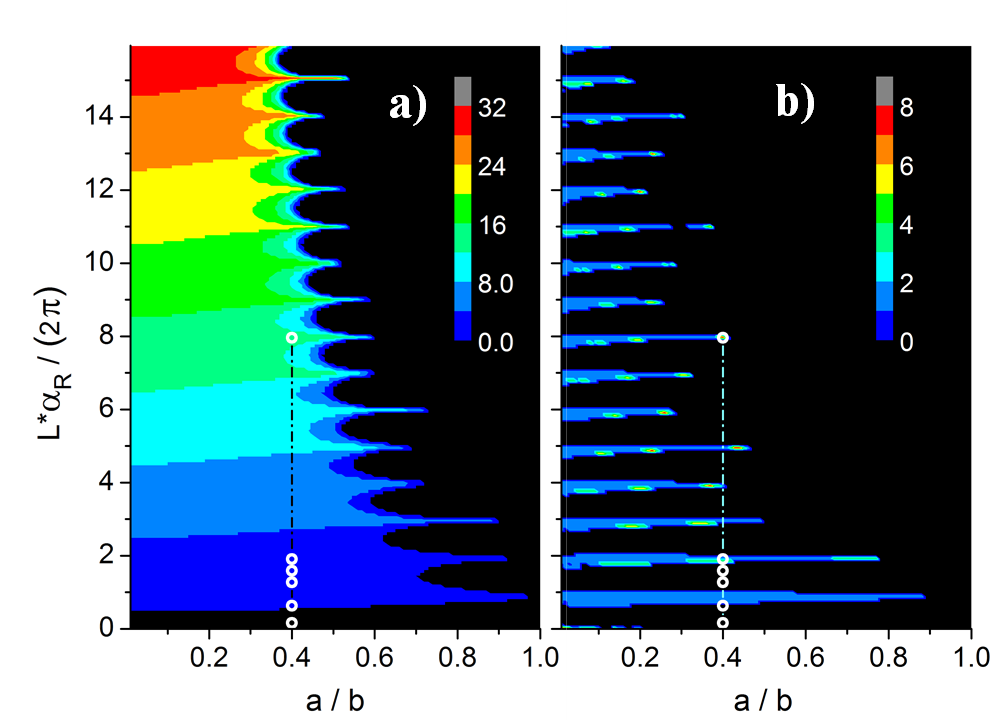}
\caption{Contour map of the winding number of the local spin orientation around (a) the normal $N$ direction in the moving frame and 
(b) the out-of-plane $z$ axis as a function 
of the ellipse ratio $a/b$ and the scaled inverse SO length $L \alpha_R/(2 \pi)$. 
The white circles at $a/b=0.4$ indicate specific points of presented spin textures (see Fig. \ref{fig:fig3}).}
\label{fig:fig2}
\end{figure}  
\begin{figure}
\includegraphics[width=0.95\columnwidth]{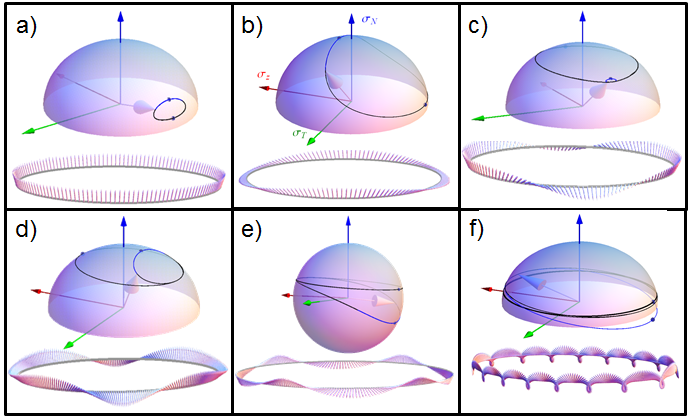}
\caption{Evolution of the spin orientation trajectories on the Bloch sphere and of the three dimensional spin textures along the ellipse shaped ring 
at a given ratio $a/b=0.4$ and for various values of the SO coupling (white circles in Fig.\ref{fig:fig2}), i.e. $L \alpha_R=1,4,8,10,12,50.1$ corresponding 
to the panels (a)-(f), respectively.
The blue portion of the trajectory stands for the part of the elliptical loop with larger curvature. 
The dots mark the positions on the loop with the maximum and minimum values of the local curvature.}
\label{fig:fig3}
\end{figure}   
{\it{Elliptical quantum rings--}}
We next investigate the spin texture in a nanostructure with non-uniform curvature by considering the 
example of a quantum ring of length $L$ with an elliptical shape and a ratio $a/b$ between the minor ($a$) and 
the major ($b$) axes of the ellipse. 
This is a paradigmatic case of a nanostructure with positive but non-uniform curvature that can be suitably enhanced (suppressed) at the 
positions nearby the poles of the major (minor) axes. In order to single out the spin textures of the eigenmodes of the Hamiltonian in Eq. \ref{eq:hamiltoniankp} 
at different regimes of Rashba SO and $a/b$ ratios, we introduce a tight-binding model obtained by discretizing Eq.~\ref{eq:hamiltoniankp} on a lattice \cite{SupMat}, 
and study the overall character of the spin textures by evaluating the local spin orientation amplitudes. 
Consistently with the gyroscope equation (Eq.~\ref{eq:gyroscope}), we find that the spin patterns as obtained from the tight-binding Hamiltonian are independent of the eigenenergies, 
and thus the profile of a single spin eigenmode is uniquely determined by $\alpha_R$ and the ratio $a/b$. 
We then characterize ``topologically" the resulting spin texture by counting the number of windings around the normal ${\hat{N}}$ and the binormal $\hat{z}$ directions the spin traces over the Frenet-Serret-Bloch sphere in a single loop. 
Fig.~\ref{fig:fig2} shows the corresponding phase diagram in the $a/b$ and $\alpha_R$ parameter space. 
We observe two distinct spin texture regimes. There is a region of the parameter space (black area), 
corresponding to configurations with strong SO or quasi-constant curvature, where the electron spin is pinned nearby the quasi-static effective field $\boldsymbol{h}_{\text{eff}}$ 
and thus does not exhibit any winding in the Frenet-Serret-Bloch sphere. 
For $a/b$ below $\sim 0.5$, {\it i.e.} for a sizable non-uniform curvature profile, the electron spin is not able to follow the 
periodic motion of the effective spin-orbit field and manifests windings both around the normal and the binormal directions. 
It follows that one can switch from a configuration without any winding to another with winding around the normal ${\hat{N}}$ and/or the binormal $\hat{z}$ directions 
through a pure geometric effect by tuning the ratio $a/b$.  

Remarkably, the onset of the spin textures with a finite number of windings can be achieved not only via a shape deformation, 
but also by an all-electrical control of the Rahsba spin-orbit inverse length $\alpha_R$. 
Below a critical threshold of the $a/b$ ratio, indeed, we find that an increase in $\alpha_R$ has a twofold effect. First, it leads to an increase in the winding number 
around the normal direction that clearly exhibits constant plateaus [Fig. \ref{fig:fig2}(a)]. 
Second, the winding number around the binormal direction has an irregular comb-like structure with a switchable on and off behavior that is 
driven by a small variation of the Rashba spin-orbit inverse length [Fig. \ref{fig:fig2}(b)]. 
\begin{figure}
\includegraphics[width=0.95\columnwidth]{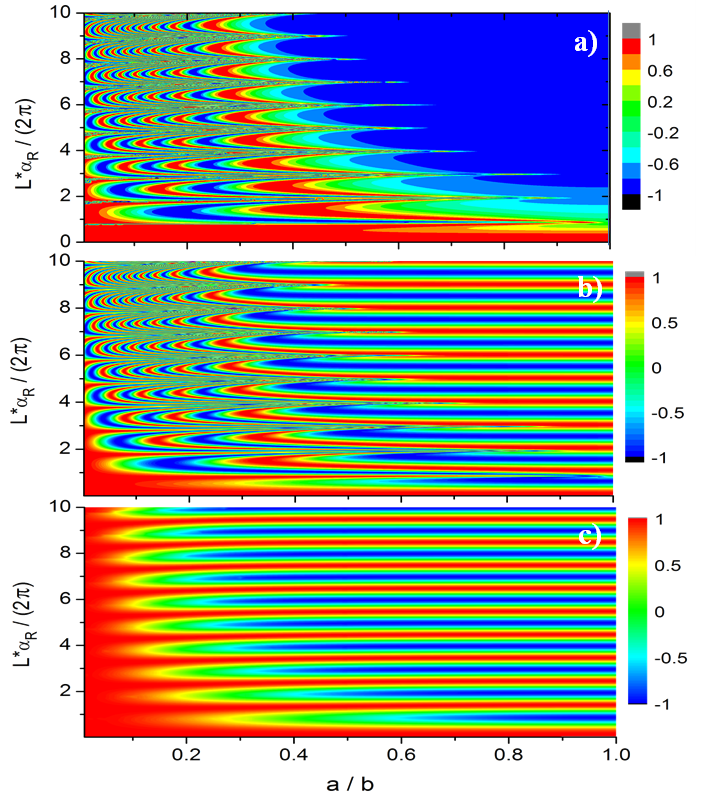}
\caption{Contour map of the cosine of the geometric phase (a), the spin component of the dynamical phase (b) and the total phase (c) contributing to the 
conductance as a function of the ellipse ratio $a/b$ and the scaled spin-orbit coupling $L \alpha_R/(2 \pi)$.}
\label{fig:fig4}
\end{figure} 
To highlight the features of the spin textures we show in Fig. \ref{fig:fig3} the 
spin trajectories in the Frenet-Serret-Bloch frame and the spin textures in the loop reference frame for few representative values of the $a/b$ and $\alpha_R$ parameters. 
Starting from a regime of SO, such as $L \alpha_R=1$, we observe in Fig. \ref{fig:fig3}(a) that the spin orientation is pinned 
around $\boldsymbol{h}_{\text{eff}}$ exhibiting a slight modulation of $\langle {\sigma}_N \rangle$ and $\langle {\sigma}_z \rangle$ and a small 
$\langle {\sigma}_T \rangle$ tangential component that changes its sign when moving along the loop. For a slightly larger SO, i.e.  $L \alpha_R=4$, 
a first transition in the winding number around the $N$ direction occurs [Fig. \ref{fig:fig3}(b)], with a spin texture that has a change of sign for $\langle {\sigma}_T \rangle$
by crossing the regions of small curvatures, close to the minor axis of the ellipse, and a slight sign change of $\langle {\sigma}_z \rangle$ nearby the 
pole of the major axis. Upon reaching $L \alpha_R=12$ the spin pattern exhibits non trivial windings around the radial direction [Fig. \ref{fig:fig3}(c)].
Hence, a further increase of $\alpha_R$ allows a transition to a configuration with windings both around $N$ and $z$ spin directions, 
with a spin texture that modulates in sign in all the three components [Fig. \ref{fig:fig3}(e)]. For intermediate $\alpha_R$ the spin texture 
has many windings along the loop [Fig. \ref{fig:fig3}(f)].
   
{\it{Geometric control of the spin transport--}}
The geometrical and the dynamical quantum phases are the key physical quantities that 
determine spin interference effects and thus the electronic transport in a cyclic quantum motion.
From the spin-torque equation (Eq.~\ref{eq:gyroscope}), we can derive
the non-adiabatic Aharonov-Anandan (AA)~\cite{AAphase} geometric phase $g_{AA}$, as well as the dynamical phase $d$ acquired 
by an electron moving along a closed loop as~\cite{SupMat}:
\begin{eqnarray}
g_{AA}&=&\pi \left(1+ \dfrac{1}{2 \pi} \int q_{N T} ds -\dfrac{1}{2 \pi} \int \dfrac{\langle {\sigma}_z \rangle}{\rho} [K(s) + q_{NT}] ds \right) \nonumber \\
d&=&d_{\sigma}+d_E
\label{eq:AAphase}
\end{eqnarray}
\noindent where $\rho$ is the local density of the spin eigenmode, $q_{N T}=\left[\langle {\sigma}_N \rangle \partial_s \langle {\sigma}_T \rangle -\langle {\sigma}_T \rangle \partial_s \langle {\sigma}_N \rangle \right] /\left[\langle {\sigma}_T \rangle^2 + 
\langle {\sigma}_N \rangle^2 \right]$,
while $d_{\sigma}=\int \alpha_R \frac{\langle {\sigma}_N \rangle}{\rho} ds$ and $d_{E}=c_0 L$ are the spin and energy dependent 
components of the dynamical phase.
From Eq.~\ref{eq:AAphase} it follows that the nature of the spin texture affects the geometric and the dynamical phase in a very different manner.
The geometric AA phase is half the solid angle spanned by the spin vector and consists of three terms:
the first $\pi$ term is the adiabatic spin geometric phase the electron would acquire when pinned at the pole of the Frenet-Serret-Bloch sphere, {\it i.e.} along the direction of the effective spin-orbit field. 
The second term corresponds, in units of $\pi$, to the winding number of the normal and tangential spin components around the out-of-plane direction. 
Due to the reflection symmetry of the ring with respect to its major axis, the latter is always an even integer, and therefore does not yields parity changes of the AA phase.
Finally, the third term yields the non-adiabatic part of the AA phase, which is directly proportional to the tilt of the electron spin in the out-of-plane $\hat{z}$ direction \cite{Frustaglia2004,Nagasawa2013}. For a circular quantum ring with $a/b \equiv 1$ this term is gradually suppressed as the Rashba spin-orbit inverse length increases, and thus the geometric phase reduces to the adiabatic $\pi$ spin geometric Berry phase. 
An analogous behavior occurs also in the case of elliptical quantum rings [the blue area in Fig. \ref{fig:fig4}(a)] in the parameter space region where the spin orientation is pinned close 
to the quasi-static effective SO field direction.  
However, as soon as the electronic spin starts to wind in the Frenet-Serret-Bloch sphere around the radial and the binormal direction, the contributions 
in the non adiabatic term yield rapid variations of the spin geometric phase as either 
the $a/b$ ratio or $\alpha_R$ are changed.
A closer inspection of the non adiabatic terms in $g_{AA}$ reveals that the contribution proportional to $\sim \langle \sigma_z \rangle \,q_{N T}$  
induces a cascade of phase slips, and therefore it represents the major player for the rapid variations of the spin geometric phase in Fig. \ref{fig:fig4}(a).
A similar behavior is also obtained for the spin dependent part of the dynamical phase that exhibits a typical oscillating behavior at a given $a/b$ as a
function of the Rashba SO in the
adiabatic regime for $a/b \sim 1$, while it turns into rapidly phase fluctuating patterns in the non adiabatic region with winding of the spin textures [Fig. \ref{fig:fig4}(b)].
 
The geometrical and dynamical phases yield the interference pattern in the 
quantum conductance. As for the case of the conventional ring~\cite{Frustaglia2004,Bercioux2005,Nagasawa2013}, 
the conductance of a single ballistic elliptical ring
symmetrically coupled to two contact leads can be obtained by means of the Landauer approach
and it is given by $G=\frac{e^2}{h}[1+\cos(d_{\sigma}+g_{AA})]$~\cite{SupMat}. Remarkably, we observe that 
in the non adiabatic regime with topological non-trivial spin textures the 
interference between the 
geometrical and dynamical phases leads to a smooth behavior of the conductance. 
The resulting pattern is marked by 
distinct {\it geometrically} driven 
channels of electronic transport with an almost constant conductance or exhibiting a
changeover from destructive to constructive interference as the $a/b$ ratio decreases (Fig. \ref{fig:fig4}(c)). 
From the analysis of the phase interference in the conductance we observe that the regular pattern of the 
non adiabatic regime mainly arises from the out-of-plane spin component 
due to a cancellation between the winding term in $g_{AA}$ and the dynamical phase $d_{\sigma}$. It is then the spatial behavior of $\langle \sigma_z \rangle$ that
dictates the conductance interference pattern by allowing a $\pi$ shift
in the region of the parameters space where both a winding around the radial and the binormal directions occurs. Such windings make the spin 
spanning a solid angle with a large amplitude on the Frenet-Serret-Bloch sphere thus reducing the adiabatic $\pi$ contribution
of the geometric phase. This result 
sets a tight connection between the changeover of the conductance and the topological character of the spin texture as one can directly observe by comparing 
the phase diagram in Fig. \ref{fig:fig2} and the patterns in Fig. \ref{fig:fig4}(c).
Evidences of a different electronic transport in the shape deformed ring can be also obtained by tuning the Rashba SO coupling through an applied gate voltage 
at a given $a/b$ ratio. The conventional oscillatory behavior in the adiabatic regime becomes damped and amplitude
modulated as one enters the region of non trivial windings (e.g. $a/b <0.5$) where the variation of the Rashba SO coupling
cannot lead to a complete destructive-constructive interference in the conductance due to the counteraction of the $q_{N T}$ term
to the dynamical spin part. 

We emphasize that our prediction can be immediately tested in the laboratory. Indeed, considering InAs quantum rings with a typical length  
$L\sim 300 $ nm, $\alpha_{SO} \sim 10^{-11}$ eV m, and the effective electron 
mass $m^* \sim 0.05 m_e$, we have  $L \alpha_{R}/(2 \pi) \sim 2$. Taking into account the gate tunability of $\alpha_{SO}$ \cite{Liang2012} and the modification of the ellipse lengths, 
one can directly access regimes of $L \alpha_{R}/(2 \pi)$ in the range of $\sim [2,10]$. By also considering that, 
apart from conventional material geometries, nanostructuring methods have 
recently achieved a level of control that even enables the synthesis of complex three-dimensional nanoarchitectures  
resembling biological structures~\cite{Xu2015},
our findings anticipate an unbound potential for new device concepts of flexible spin-orbitronics where the electron spin 
and the electronic transport are directly controlled by the system geometry. 

{\bf{Acknowledgements}}  We acknowledge the financial support of the Future and Emerging Technologies (FET) programme under FET-Open grant number: 618083 (CNTQC). 
CO thanks the Deutsche Forschungsgemeinschaft (grant No. OR 404/1-1) for support.

\section{Appendix}

In this section we present the derivation of the spin-torque equation for the
spatial evolution of the local spin orientation in a generic planarly curved nanowire 
with Rashba spin-orbit coupling. We also demonstrate how
the spin velocity vector field behaves around the positions corresponding to the locally vanishing
effective spin-orbit torque. Then, we present the expressions for the geometric and dynamical phases acquired by 
an electron going around a loop having a non trivial geometric curvature and Rashba spin-orbit coupling and derive the 
conductance for the quantum loop. Finally, we provide
the details for the continuum-to-lattice mapping of the Hamiltonian for a 
curved nanowire and the methodology applied to determine the phase diagrams.  
\subsection {Gyroscope equation for the local spin orientation}  

We consider a generic planarly curved 
one-dimensional nanostructure
in the presence of Rashba spin-orbit coupling  
described by the Hamiltonian ${\cal H}_{\bf k \cdot p}$. From the structure of
${\cal H}_{\bf k \cdot p}$ it follows that:
\begin{eqnarray*}
{\cal H}_{\bf k \cdot p}=H^2_l-\frac{\alpha_{SO}^2}{4 \gamma} \sigma_0
\end{eqnarray*}
\noindent with $\sigma_0$ being the identity matrix, $\gamma=\frac{\hbar^2}{2 m^*}$ and $H_l$ reads:
\begin{eqnarray*}
H_l=\left( i \sqrt{\gamma} \partial_s+ \frac{\alpha_{SO}}{2 \sqrt{\gamma}} \sigma_N(s) \right)
\end{eqnarray*}
Hence, $H_l$ and ${\cal H}_{\bf k \cdot p}$ have common eigenstates with an eigenvalue relation given by $E_{\bf k \cdot p}=E^2_l-\frac{\alpha_{SO}^2}{4 \gamma}$.
Let us introduce the local spin orientation for a given spin eigenmode $|\Psi _{E}\rangle$ as the corresponding expectation value
of the spin operators in the Frenet-Serret reference frame, i.e.    
$\langle \boldsymbol{\sigma} \rangle=\{\langle {\sigma}_T \rangle,\langle {\sigma}_N \rangle,\langle {\sigma}_z \rangle \}$.
To determine the equations for the spatial derivative of the local spin components it is convenient to use the 
relation $H_l |\Psi _{E}\rangle =E_l |\Psi _{E}\rangle$ 
for the spin eigenmode wavefunction in such a way to single out the spatial 
derivative as
\begin{eqnarray}
i \partial_s |\Psi _{E}\rangle &&= G(s) |\Psi _{E}\rangle \label{evolve} \\
i \partial_s \langle \Psi _{E}| &&= - \langle \Psi _{E}| G(s) \nonumber
\end{eqnarray}
with $G(s)=- \left( \alpha_R \sigma_N(s)+ c_0\,\sigma_0\right)$, $c_0^2=\frac{2 m^{*} E}{\hbar^2}+\alpha_{R}^2$ relates to the eigenergy $E$, 
and 
$\alpha_R=\frac{\alpha_{SO}}{2\gamma}$. 
From the Frenet-Serret equations for the normal and tangential directions to the curved 
nanostructure,
$\partial_s \hat{\cal N}(s) = K(s) \hat{\cal T}(s)$ and
$\partial_s \hat{\cal T}(s) = -K(s) \hat{\cal N}(s)$,  
it follows
that $\partial_s {\sigma}_N(s)=K(s) {\sigma}_T(s)$ and $\partial_s {\sigma}_T(s)=-K(s) {\sigma}_N(s)$. Hence, 
taking into account the Eqs. \ref{evolve} for the spatial evolution of the eigenmode, one can obtain the general   
expression for the spatial derivative of the expectation value of the spin components
\begin{eqnarray}
\partial_s \langle \boldsymbol{\sigma} \rangle = i\langle [G, \boldsymbol{\sigma}]\rangle + \langle \partial_s \boldsymbol{\sigma} \rangle \,
\label{eq:deriv}
\end{eqnarray}
\noindent with $[A,B]$ indicating the commutator of $A$ and $B$. 

At this stage, by combining the Frenet-Serret equations for the spin operators and the Eq. \ref{eq:deriv}, and considering that $[G,{\sigma}_N]=0$,    
$[G,{\sigma}_T]=2 i \alpha_R {\sigma}_z$, $[G,{\sigma}_z]=-2 i \alpha_R {\sigma}_T$, and $\partial_s \sigma_z=0$,  it follows:
\begin{eqnarray}
\partial_s \langle \sigma_N \rangle &=& K(s) \langle \sigma_T \rangle \nonumber \\
\partial_s \langle \sigma_T \rangle &=& -2 \alpha_R \langle \sigma_z \rangle - K(s) \langle \sigma_N \rangle \nonumber \nonumber \\
\partial_s \langle \sigma_z \rangle &=& 2 \alpha_R \langle \sigma_T \rangle 
\label{eq:gyro}
\end{eqnarray}
The previous relations for  $\partial_s \langle \boldsymbol{\sigma} \rangle$ can be rewritten in a compact gyroscope-like form by introducing the effective spatial 
dependent field $\boldsymbol{h}_{\text{eff}}=\{0,-2 \alpha_R, K(s)\}$ in the basis of
the spin components $\langle \boldsymbol{\sigma} \rangle=\{\langle {\sigma}_T \rangle,\langle {\sigma}_N \rangle,\langle {\sigma}_z \rangle \}$ as
\begin{eqnarray*}
\partial_s \langle \boldsymbol{\sigma} \rangle= \boldsymbol{h}_{\text{eff}} \times \langle \boldsymbol{\sigma} \rangle \,. \label{eq:gyro}
\end{eqnarray*}

For completeness, we notice that from Eq. \ref{eq:deriv} one can also obtain that the local electron density $\rho(s)=\langle \Psi|\Psi\rangle$ is spatially constant 
as $\partial_s \langle \sigma_0 \rangle =0$. 
Moreover, since the spatial derivative of the spin vector is perpendicular to $\boldsymbol{\sigma}$ it follows that 
the amplitude of the local spin component $\langle \boldsymbol{\sigma} \rangle^2$ is spatially uniform, i.e. 
$\partial_s \left(\langle \boldsymbol{\sigma}\rangle \cdot \langle \boldsymbol{\sigma}\rangle \right) = 0$.

\subsection{Character of the spin-orientation velocity field}  
We have shown that the amplitude of the spin is constant along the profile of the nanostructure, and 
thus the spin trajectory can be analyzed by 
introducing a Bloch sphere in the Frenet-Serret reference
frame, where a point on the sphere identifies the spin orientation at a given position $s^*$ through the angles $\{\theta(s^*),\varphi(s^*)\}$ (see Fig. 1 in the 
main text of the manuscript). In spherical coordinates the spin components can be expressed as 
\begin{eqnarray*}
\langle \sigma _{N}(s)\rangle &=&\sigma \cos [\theta ] \\
\langle \sigma _{z}(s)\rangle &=&\sigma \sin [\theta ]\cos [\varphi ] \\
\langle \sigma _{T}(s)\rangle &=&\sigma \sin [\theta ]\sin [\varphi ] \,.
\end{eqnarray*}
\noindent With this, the Eqs. (3) reduce to two independent equations for the derivative of the coordinates $\{\theta(s),\varphi(s)\}$ 
\begin{eqnarray*}
\overset{\cdot }{\theta } &=&-K(s)\sin[\varphi ] \\
\overset{\cdot }{\varphi } &=&-2\alpha _{R}-K(s)\cos[\varphi ]\frac{1}{\tan
[\theta ]} \,.
\end{eqnarray*}
In order to analyze the character of the possible stable points for the spin orientation trajectories 
we search for solutions where the polar $\overset{\cdot }{\theta}$ and azimuthal $\overset{\cdot }{\varphi}$ 
velocities are vanishing. Assuming that the curvature is non singular along the nanostructure, the
polar and the azimuthal velocities are zero at the points $P_{1,2}$ such as $\varphi=\overline{\varphi }_{1,2} =$0 or $\pi$ 
and $\theta=\overline{\theta }_{1,2}(s) =arccot [\pm \frac{2\alpha _{R}}{K(s)}]$. It is worth pointing out that the 
vanishing of the effective torque occurs for points that lie in the normal-binormal plane for the spin components independently on the geometric 
properties of the nanostructure.
On the other hand, the critical $\overline{\theta }_{1,2}(s)$ evolve along the Frenet-Serre-Bloch sphere when changing the position $s$ on the
curved profile. One can easily obtain the cinematic parameters (velocity and acceleration) 
of these points on the sphere by evaluating the first and second derivatives as 
\begin{eqnarray*}
v_{\overline{\theta }_{1,2}}(s)&=&\pm \left[ K^{'}(s)\right] 
\frac{2 \alpha _{R}}{\left(4 \alpha _{R}^{2}+K(s)^{2}\right) } \\
a_{\overline{\theta }_{1,2}}(s)&=& \pm \frac{2 \alpha_{R} [\left(4 \alpha _{R}^{2}+K(s)^{2}\right)K^{''}(s)-2 K(s) K^{'}(s)^2 }{\left(4 \alpha _{R}^{2}+K(s)^{2}\right)^2} \,.
\end{eqnarray*}%

Then, the evolution of the $\overline{\theta }_{1,2}(s)$ is
strongly 
interconnected to the geometry of the nanosystem
not only through the strength of the curvature but also through the 
derivatives of the curvature. 
This provides
a direct connection between the geometric curvature of the
structure and the cinematic of the points where the velocities of the
local spin components vanish. To further understand the character of the spin vector flow
we determine the Jacobian $J$ of the first derivatives of the velocities at the points $P_{1,2}(s)$:
\begin{equation*}
J=%
\begin{pmatrix}
0 & \pm K(s) \\ 
\mp K(s)\ \frac{K(s)^{2}}{\left( \alpha _{R}^{2}+K(s)^{2}\right) } & 0\ 
\end{pmatrix} \,.
\end{equation*}%
The eigenvalues of the matrix $J$ provide informations
on the character
of the solutions around the points $P_{1,2}$. We find that the eigenvalues $E_{J}$, 
independently on the positions 1,2, are given by
\begin{equation*}
E_{J}=i \frac{K(s)^{2}}{\sqrt{4 \alpha _{R}^{2}+K(s)^{2}}} \left\{ -1 ,1 \right\} \,.
\end{equation*}%
The eigenvalues are imaginary for any value of the
curvature and the renormalized Rashba coupling $\alpha_{R}$ and thus the solutions for $%
\theta $ and $\varphi$ to the linear order are oscillating. This result implies that around the
points $P_{1,2}(s)$ the trajectories of the velocity fields can form
closed loops, namely they have a vortex like profile. When the curvature
varies along the profile of the one-dimensional nanostructure the vortex moves back and forth
along the Frenet-Serret-Bloch sphere from the north(south) pole to the equator.

\subsection{Aharonov-Anandan geometric and dynamical phases}  

In order to determine the geometric phase acquired by an electron moving along a closed loop, we follow the original approach 
proposed by Aharonov-Anandan for any cyclic evolution of a quantum system \cite{AAphase}.  
A spin eigenmode $|\Psi (s)\rangle$ of ${\cal H}_{\bf k \cdot p}$ evolves in space according to the Eq. \ref{evolve}. 
Now, if we assume that the evolution is along a closed curve
of length $L$, the wave functions $|\Psi (s)\rangle$ at the initial and final positions of the loop are related by a phase factor $e^{i \beta}$, with $\beta$ real, such as 
$|\Psi (L)\rangle=e^{i \beta} |\Psi (0)\rangle$. 
Then, we can define $|\tilde{\Psi} (s)\rangle =e^{-i b(s)} |\Psi (s)\rangle$ in such a way that $b(L)-b(0)=\beta$. It immediately follows that  
$|\tilde{\Psi} (L)\rangle=|\tilde{\Psi} (0)\rangle$ and from Eq. \ref{evolve} that
\begin{eqnarray*}
-\partial_s b(s)= \frac{\langle \Psi| G(s) |\Psi\rangle}{\langle \Psi| \Psi \rangle}-\frac{\langle \tilde{\Psi}| i\partial_s |\tilde{\Psi}\rangle}{\langle \Psi|\Psi \rangle} \,.
\end{eqnarray*}
The total phase $\beta$ acquired by the charge carrier along the loop is given by the sum of a geometric $g_{AA}$ and dynamical part $d$ \cite{AAphase} as
\begin{eqnarray}
g_{AA}= \int_{0}^{L} \frac{\langle \tilde{\Psi}| i\partial_s |\tilde{\Psi}\rangle}{\langle \Psi|\Psi \rangle} ds \label{eq:gAA} \\ 
d=-\int_{0}^{L} \frac{\langle \Psi| G(s) |\Psi\rangle}{\langle \Psi| \Psi \rangle} ds\,.
\end{eqnarray}

Now, we are interested in obtaining $g_{AA}$ for 
a generic closed curve. In doing this, we aim to obtain 
an expression that directly links the geometric phase
to the local spin orientation of the evolving quantum state. 
The wave-function $|\Psi (s)\rangle$ can be generally expressed 
in the form
\begin{equation*}
|\Psi _{E} \rangle=\left( 
\begin{array}{c}
\exp[-i f(s)/2]\, \exp[i \theta_{\Uparrow}(s)] A_{\Uparrow}(s) \  \\ 
\exp[i f(s)/2]\, \exp[i \theta_{\Downarrow}(s)] A_{\Downarrow}(s)%
\end{array}%
\right)
\end{equation*}%
\noindent where $f(s)=\int_{0}^{s} K(\bar{s}) d\bar{s}$, and $\{A_{\Uparrow},A_{\Downarrow}\}$ are real. This structure for $|\Psi(s) \rangle$ is 
convenient because the expectation values of  
the local spin $\langle \boldsymbol{\sigma} \rangle$ in the Frenet-Serret reference frame is related to the components of the wave-function by the following relations:
\begin{eqnarray}
\tan[\theta_{\Uparrow}-\theta_{\Downarrow}]=\frac{\langle {\sigma}_T \rangle}{\langle {\sigma}_N \rangle} \label{wind} \\
A_{\Uparrow}^2-A_{\Downarrow}^2=\langle {\sigma}_z \rangle \label{sz}
\end{eqnarray}
Furthermore, the local density is given by $\rho(s)=A_{\Uparrow}^2+A_{\Downarrow}^2$ and $\int_{0}^{L} K(\bar{s}) d\bar{s}=2 \pi$. 
The phase difference $(\theta_{\Uparrow}-\theta_{\Downarrow})$ acquires
a shift 
$2\pi W_{N T}$ that is multiple of 2$\pi$ going around the loop from 0 to $L$, with $W_{N\,T}$ 
given by
\begin{eqnarray*}
W_{N\,T}=\int_{0}^{L} q_{NT} ds \,.
\end{eqnarray*}
with $q_{NT}=\frac{\left[\langle {\sigma}_N \rangle \partial_s \langle {\sigma}_T \rangle -\langle {\sigma}_T \rangle \partial_s \langle {\sigma}_N \rangle \right]}{\left[\langle {\sigma}_T \rangle^2 + 
\langle {\sigma}_N \rangle^2 \right]}$ corresponding to the winding of the normal and tangential spin components around the binormal direction. Hence, one can show that
\begin{equation*}
|\tilde{\Psi} (s) \rangle= \left( 
\begin{array}{c}
A_{\Uparrow}(s) \  \\ 
\exp[i f(s)]\, \exp[-i (\theta_{\Uparrow}(s)-\theta_{\Downarrow}(s))] A_{\Downarrow}(s)%
\end{array}%
\right) \,.
\end{equation*}%
Now, it follows that the geometric phase in Eq. \ref{eq:gAA} can be directly obtained by expanding the derivative of $|\tilde{\Psi} (s) \rangle$ and
using the relations between the components and the spin amplitudes. The resulting expression is
\begin{eqnarray*}
g_{AA}=\pi \left(1+ \dfrac{1}{2 \pi} \int q_{NT} ds -\dfrac{1}{2 \pi} \int \dfrac{\langle {\sigma}_z \rangle}{\rho} [K(s) + q_{NT}] ds \right) \,.
\end{eqnarray*} 

Finally, by taking the expression of $G(s)$ we can immediately obtain the dynamical phase as:
\begin{equation}
d=d_{\sigma}+d_E
\end{equation}
\noindent 
$d_{\sigma}=\int \alpha_R \frac{\langle {\sigma}_N \rangle}{\rho} ds$ and $d_{E}=c_0 L$ are its spin and energy dependent 
components.

\subsection{Conductance of an elliptical quantum loop}

In this section we derive the expression of the quantum conductance for the elliptical ring 
assuming that it is symmetrically coupled to contact leads in the limit of low bias applied voltage.
In order to obtain the conductance we employ the Landauer formula and follow the approaches of Refs. \cite{Frustaglia2004,Bercioux2005} assuming fully transparent
contacts between the leads and the ring, and
neglecting backscattering effects that lead to resonances. 
The performed analysis for the one-dimensional elliptical ring is based on semiclassical 
method that gives direct access to the local spin dynamics and the geometric phases in the ballistic regime. 
Such approach has been demonstrated to successfully reproduce the key features of the 
electronic transport for a realistic quasi two-dimensional ring geometry \cite{Frustaglia2004,Saarikoski2015}.
We assume two possible and equally probable paths to transmit the spin carriers along the upper and lower arms of the loop.
The resulting transmission amplitude matrix is given by $\Gamma=\Gamma_{u}+\Gamma_{l}$ where $\Gamma_{u}$ and $\Gamma_{l}$ are the matrices 
associated to the transmission on the upper and lower arm of the ring,
respectively. 
These transmission amplitudes can be obtained by approximating the loop by a polygon with
a large number of vertices as applied to the case of a circular ring \cite{Bercioux2005} .
As a consequence of the Eq. \ref{evolve}, the evolution of the spin vector along each side of the polygon 
can be described by a spin rotation operator that gives the spin precession around the effective spin-orbit field,
\begin{eqnarray*}
R_{i,j}=e^{-i \sigma_N(s_{i,j}) \alpha_R l_{i,j}} 
\end{eqnarray*}
\noindent 
where $\sigma_N(s_{i,j})$ is the spin component that is normal to the vector connecting the vertices $i$ and $j$
and evaluated at the average 
position $s_{i,j}$ between them. $l_{i,j}$ is the distance between the vertices.
Assuming that the paths along the upper and lower arm of the loop are decomposed into $N$ segments and that the
leads are connected at the vertices $1$ and $N/2$,
the quantum amplitude for the transmission can be obtained as the sum of the successive application
of the operator $R_{i,j}$ for each channel:
\begin{eqnarray}
\Gamma_{u,N}&=&R_{N/2,N/2-1}\cdot \cdot \cdot R_{3,2}\cdot R_{2,1} \nonumber \\
\Gamma_{l,N}&=&R_{N/2,N/2-1}\cdot \cdot \cdot R_{N-1,N}\cdot R_{N,1} \nonumber \\
\Gamma_{N} &=& \Gamma_{u,N}+\Gamma_{l,N}\,.
\end{eqnarray}
The transmission coefficient is then given by 
\begin{eqnarray*}
T_{N}=Tr[\Gamma_{N} \Gamma^{\dagger}_{N}]
\end{eqnarray*}
The zero temperature conductance at the lowest order of approximation for each spin channel can expressed by:
\begin{equation*}
G_N=
\frac{e^{2}}{2 h } T_{N} \, ,
\end{equation*}
\noindent with $e$ being the electron charge and $h$ the Planck constant.
In order to obtain the amplitude of the conductance for the elliptical quantum loop, we 
perform the limit $N\rightarrow \infty$. We notice that the key contributions to $T_N$ arise 
from the terms $\Gamma_{ul}=\Gamma_{u,N} \Gamma^{\dagger}_{l,N}$ and  $\Gamma_{lu}=\Gamma^{\dagger}_{ul}$.
In the limit of large $N$ the two-paths interference operator is given by $\Gamma_{ul}=e^{-i \alpha_R \int^{L}_{0} \sigma_N(s) ds}$.
By using the relation  $\alpha_R \sigma_N(s)=-G(s)-c_0$, and observing that the operator $e^{-i\int^{L}_{0} G(s) ds}$ has 
eigenvalue $e^{-i \phi}$ with $\phi=g_{AA}+c_0 L+d_\sigma$, one can immediately determine the trace of $T$ and consequently 
the linear conductance: 
\begin{equation}
G=\frac{e^{2}}{h }\left\{ 1+\cos \left[g_{AA}+d_{\sigma
} \right] \right\} .
\end{equation}%

We point out that the  expression derived above for the conductance
holds for any closed curved
with a generic 
curvature profile $K(s)$.

\subsection{Tight-binding approximation}

\noindent Here, we present the mapping of the continuum Hamiltonian on the lattice of a planarly curved one-dimensional nanostructure and
the methods for the computation of the spin texture. 
The phase diagrams corresponding to the spin textures and the geometric phase 
of the spin eigenmode for the elliptical shaped ring have been obtained by solving the Rashba model Hamiltonian on an effective lattice system. 
By employing a 
conventional discretization procedure from the derivative in the continuum to the finite differences in the lattice, 
one can show that the Hamiltonian ${\cal H}_{\bf k \cdot p}$ can be mapped into 
an effective lattice model \cite{Gentile2015} that reads
\begin{equation}
{\cal H}=\sum_{j} \sum_{\sigma,\sigma'=\uparrow,\downarrow}
c^{\dag}_{j,\sigma}(t \ \delta_{\sigma, \sigma'} +  \hat{\alpha}_{j,j+1}^{\sigma,\sigma'})c_{j+1,\sigma'} +{\it h.c} , 
\label{eq:hamiltonian}
\end{equation}
\noindent where $c_{j,\sigma}^{\dag}, c_{j,\sigma}$ are operators creating and annihilating, respectively, an electron at the $j$-th site with spin projection $\sigma= \uparrow, \downarrow$ 
along the $z-$axis, $t=\frac{\hbar^2}{2 m^* d^2}$ is the hopping amplitude between nearest-neighbor sites with $d$ being the distance between two neighboring sites. 
We assume in the computation that the curvature does not change the hopping amplitude for nearest-neighbor distances. This aspect does not influence the qualitative outcome of the results.  
The local Rashba spin-orbit coupling connecting nearest-neighbour sites is $\hat{\alpha}_{j,j+1}$ and can be written in terms of the set of Pauli matrices ${\boldsymbol \tau}$ as
$\hat{\alpha}_{j,j+1}= i \frac{\alpha_{SO}}{4 d} \, \left[ \tau_x \, g_j^{x} + \, \tau_y \, g_j^y \right]$ . 
In the equation above, $g_j^{x}= \cos{f(s_j)} + \cos{f(s_{j+1})}$ and $g_j^{y}= \sin{f(s_j)} + \sin{f(s_{j+1})}$, which are determined by the position of the sites along 
the curved line
and the specific geometrical shape encoded in the function $f(s)$ via the curvature $K(s)$. The model Hamiltonian has been then 
diagonalized assuming that the closed curve
has a given length $L$ and is composed of a given number of sites $N$. The computation of the expectation values of the 
local spin components
of the eigenmodes has been performed for a number of sites $N$ varying from 500 to 1000. The results do not depend qualitatively from the number $N$ and are 
fully consistent with the prediction of the gyroscope equation obtained for the continuum model.


\begin{thebibliography}{99}
\bibitem{SupMat} In the appendix we provide details about the derivation of the gyroscope-like equation for the
spatial evolution of the local spin orientation in a generic planarly curved nanowire 
with Rashba spin-orbit coupling. We demonstrate how
the spin velocity vector field behaves around the positions corresponding to the locally vanishing
effective spin-orbit torque. Then, we derive the expression for the geometric phase acquired by 
an electron going around a loop having a non trivial geometric curvature and Rashba spin-orbit coupling. 
We derive the expression for the quantum conductance of an elliptical ring symmetrically coupled to two contact leads.
Finally, we provide the details for the continuum-to-lattice mapping of the Hamiltonian for a 
curved nanowire and the methodology applied to determine the phase diagrams.
\end{thebibliography}
\end{document}